\newcommand{\rem}[1]{}
\newcommand{\hbo}{\hbar\omega}

\documentclass[pra,twocolumn,showpacs,preprintnumbers,amsmath,amssymb,floatfix]{revtex4}
\usepackage{hyperref}
\usepackage{epsfig}
\begin{document}

\title{
Low-energy excitations of a linearly Jahn-Teller coupled orbital quintet.
}

\author{Nicola Manini}
\email{nicola.manini@mi.infm.it}
\affiliation{Dipartimento di Fisica, Universit\`a di Milano,
Via Celoria 16, 20133 Milano, Italy}
\affiliation{INFM, Unit\`a di Milano, Milano, Italy}
\affiliation{International School for Advanced Studies (SISSA),
Via Beirut 4, 34014 Trieste, Italy}

\date{September 30, 2004}

\begin{abstract}

The low-energy spectra of the single-mode $h\otimes (G+H)$ linear
Jahn-Teller model is studied by means of exact diagonalization.
Both eigenenergies and photoemission spectral intensities are computed.
These spectra are useful to understand the vibronic dynamics of icosahedral
clusters with partly filled orbital quintet molecular shells, for example
C$_{60}$ positive ions.
\end{abstract}


\maketitle

\section{Introduction}

In icosahedral molecules and ions, there occurs the possibility of a
partial occupancy of a quintet of orbitally degenerate electronic levels,
of $h$ symmetry.
A notable example is that of positive C$_{60}$ ions, with one (or several)
holes in a $h_u$ molecular orbital \cite{Lueders03,Bohme99}.
At the adiabatic level, partly occupied quintet states are unstable against
molecular distortions involving the vibrational modes of the molecule.
Distortions along nondegenerate $A$ modes do not involve symmetry
reduction, are therefore not of Jahn-Teller (JT) type, and can be trivially
treated as phonon shifts: thus we will ignore $A$ modes altogether.
Distortions involving fourfold-degenerate $G$ and fivefold-degenerate $H$
modes reduce the molecular symmetry, and are responsible for the leading
(linear) contribution to the splitting of the fivefold molecular level.
Other ($T_1$ and $T_2$) modes are involved in higher-order couplings, which
become dominating only for substantially large distortions.
Under specific circumstances (null or very strong linear coupling)
the inclusion of higher-order couplings could become essential.
However, the present work follows the standard practice of neglecting all
higher-then-linear coupling terms and treating the phonons as harmonic
oscillators \cite{Ceulemans,CeulemansII}.
This approximation provides a simplified model meant to capture the basic
shape of the five adiabatic potential energy surfaces (APES) in a
neighborhood of their high-symmetry conical intersection, hopefully
extending out to include the JT minima.
The advantage of the idealized linear model is that it involves a minimal
number of parameters, and that it can therefore be studied in detail,
drawing fairly general conclusions, applicable to a wide enough range of
realistic systems.

Indeed, besides C$_{60}$, orbital quintets could be relevant for the
spectroscopy of other icosahedral molecules and clusters among those
synthesized in recent years, including smaller and higher fullerenes such
as C$_{20}$ and C$_{80}$, B$_{12}$ or B$_{84}$ clusters, B$_{12}$H$_{12}$,
the smallest icosahedral systems Si$_{13}$, Na$_{13}$, or Mg$_{13}$, and
icosahedral structures with 13, 19, 55, and 147 atoms in xenon clusters.
With different coupling patterns possibly realized in those different
systems, it is of clear interest to investigate how the vibronic spectrum
changes as a function of electron-vibration coupling.
For simplicity, here we restrict ourselves to a single mode (of either $G$
or $H$ type) coupled to the orbital quintet.

Several works \cite{Ceulemans,CeulemansII,Szopa} attach the JT problem of
the orbital quintet in the linear approximation and for a single mode.
These early works are mainly studies of the APES, and in particular of its
minima, with the phonon coordinates treated as classical variables.
Other papers deal with the full quantum mechanical problem, and provide
either approximate \cite{Moate96,Moate97,Polinger03} or exact
\cite{Delos96,hbyh} solutions, sometimes for the many-modes
case \cite{Manini03,ManyModes}.
The quantum mechanical dynamics of the entangled electron-phonon system is
affected by an electronic geometric phase, whose presence/absence has
relevant consequences, in particular, for the ground-state (GS) symmetry
\cite{AMT,Delos96,Paris97,Polinger03,Koizumi99,Lijnen03}, thus for magnetic
resonance spectroscopies \cite{Abragam}.
In particular, Ref.~\onlinecite{hbyh} provides a detailed study of the
evolution of the tunneling splitting through the range of coupling
parameters for an $H$ mode\footnote{
All quoted numerical values of $g$ in Ref.~\onlinecite{hbyh} are incorrect
and should be multiplied by $2^{1/2}$.
}.

Unfortunately, to date, no experimental determination of the tunneling
splitting in any icosahedral JT system is available.
In contrast, spectroscopic techniques can and do provide direct measurements
of the JT vibronic spectrum.
For example, molecular photoemission
\cite{Bruhwiler97,Canton02} was recently shown \cite{Manini03} to provide
relatively detailed information on the low-lying part of the vibronic
spectrum of C$_{60}^+$, especially in a specific symmetry sector
\cite{ManiniComm03}.
A clean theoretical understanding of the general features of this vibronic
spectrum is therefore desirable.

The present work provides precisely an exact determination of the low-lying
vibronic levels of the single-mode linear model, those usually of most
direct experimental access.
The spectrum is studied through a wide range of coupling parameters, from
weak to strong coupling.
In practice, the reported results are mainly relevant for weak to
intermediate coupling, since, as noted above, the linear JT model is little
more than a mathematical curiosity in the strong-coupling limit, where
higher-than-linear powers of the distortion become dominant.
The vibronic spectra presented in this work illustrate several interesting
quantum phenomena characteristic of dynamical JT.
Several numerical energies reported in Appendix \ref{app} can be used
as benchmarks of the accuracy of future calculations based on approximate
approaches, for example of the type of Refs.~\onlinecite{Moate96,Dunn03}.

\section{Model and calculation}

The model \cite{CeulemansII,hbyh,Moate96} describing the JT coupling of
the orbitally degenerate $h$ quintet with the molecular vibrations of
symmetry $\Lambda = G,\,H$ is conveniently formulated as follows
\cite{hbyh,Manini01,Lueders02}:
\begin{eqnarray}
\label{modelhamiltonian}
\hat{H} &=& \hat{H}_0 + \hat{H}_{\rm vib} + 
\hat{H}_{\rm e-v} \,,\\
\hat{H}_0     &=& \epsilon \, \sum_{m \sigma}  
\hat{c}_{m \sigma}^\dagger \hat{c}_{m \sigma} \,, \\
\label{vib-hamiltonian}
\hat{H}_{\rm vib} &=& \frac 12 \hbo \sum_{\mu}
\left(\hat{P}_{\mu}^2+\hat{Q}_{\mu}^2\right) \,, \\
\label{JT-hamiltonian}
\hat{H}_{\rm e-v} &=&  \hbo
\, k^\Lambda
\sum_{{r \sigma \, \mu m m'}}
g^{(r)}_{\Lambda}
C^{r \Lambda \mu}_{m \; -m'} \,
\hat{Q}_{\mu} \,
\hat{c}^\dagger_{m\sigma} \hat{c}_{m' \sigma } \,.
\end{eqnarray}
%
%
Here $m$, $\mu$ label components within the degenerate multiplets, for
example according to the $C_5$ character in the ${\cal I}\supset D_5\supset
C_5$ group chain \cite{hbyh,Butler81},
$C^{r \Lambda \mu}_{m m'}$ are icosahedral Clebsch-Gordan coefficients
\cite{Butler81} which couple two $h$ tensor operators and a $\Lambda$
tensor operator to a global scalar $A$ operator.
$\hat{Q}_{\mu}
$ are the dimensionless normal-mode vibrational coordinates (in units of the
natural length scale of the harmonic oscillator), and $\hat{P}_{\mu}$ the
corresponding conjugate momenta.
The multiplicity $r=1,2$, needed for $\Lambda = H$ vibrations only, labels
the two separate kinds of coupling allowed under the {\em same} symmetry
\cite{Manini01,Butler81}: it represents the double occurrence of the $H$
representation in the direct product $h\times h$ (the icosahedral group is
not simply reducible).
Numerical factors 
$k^{G}=\frac 14 {5^{\frac 12}}$, $k^{H}=\frac 12$ (which could otherwise be
re-absorbed into the definition of $g_{\Lambda}$) are included to make
contact with the notation of Ref.~\onlinecite{Manini01}.

Without any loss of generality, the energy position $\epsilon$ of the $h$
quintet will be taken as the zero of energy.
In this single-mode problem, all energies are naturally measured in units
of the vibrational quantum $\hbo$.
The second characteristic energy scale $g_{\Lambda}^2 \hbo$ (the JT energy)
is smaller/larger than $\hbo$ for small/large $g_{\Lambda}$ (weak/strong
coupling).
The weak-coupling regime is characterized by rapid tunneling among shallow
clustered potential wells and strong non-adiabaticity associated to the
vicinity to the conical intersection of the five Born-Oppenheimer potential
sheets; the level structure is perturbatively related to the harmonic
spectrum of $\hat{H}_{\rm vib}$.
As the coupling grows to intermediate and strong, the JT wells deepen and
move away from the conical intersection and from each other: tunneling
decreases and the vibronic spectrum acquires a characteristically intricate
structure, which simplifies again in the strong-coupling limit, where
tunneling is exponentially suppressed and semiclassical considerations
apply
\cite{Polinger03}.

The adiabatic approximation, where the vibrational kinetic energy is
neglected, provides the main (static JT) features of the strong-coupling
limit.
The distortion operators $\hat{Q}_{\mu}$ are treated as classical
coordinates, and optimal distortions are obtained by minimizing the lowest
APES \cite{Moate96,Moate97}, for example by means of the \"Opik-Pryce
method \cite{OP}.
These optimal distortions (JT wells, or minima) are associated to specific
splitting patterns of the electronic quintet.
In particular, $G$ modes produce a level pattern of the type
$(-8,-3,-3,7,7)$, in correspondence to 10 JT wells of $D_3$ residual
symmetry.

The situation is slightly more intricate for an $H$ mode
\cite{CeulemansII,hbyh,Moate96}, due to the multiplicity $r=1,\,2$.
The coupling scheme, associated to the $r\!=\!2$ Clebsch-Gordan
coefficients, produces the splitting pattern $(-4,-1,-1,3,3)$ in
correspondence to 10 $D_3$ minima, like the $G$ modes.
The $r\!=\!1$ scheme produces a splitting pattern of the type
$(-4,1,1,1,1)$, in correspondence to 6 JT minima of $D_5$ symmetry.

In practice, the two-scheme coupling for the $H$ modes is a convenient
theoretical idealization, since any $H$ mode of a real molecule is
associated to some amount of coupling of both $r\!=\!1$ and $r\!=\!2$ kind.
This admixture is conveniently characterized by introducing {\em two}
coupling constants for each $H$ mode, rather than one.
The two dimensionless linear coupling parameters $g_H^{(1)}$, $g_H^{(2)}$
can be expressed in polar coordinates \cite{hbyh,Manini01} as
\begin{equation}
g_H^{(1)} = g \, \cos\alpha  \,, \qquad
g_H^{(2)} = g \, \sin\alpha  \,.
\end{equation}
In summary, besides its frequency $\omega$, each $H$ vibrational mode is
characterized by a pair ($g_H^{(1)}$, $g_H^{(2)}$) of linear couplings, or
equivalently a global coupling intensity $g$ plus a mixing angle $\alpha$.

\begin{figure}
\centerline{
\epsfig{file=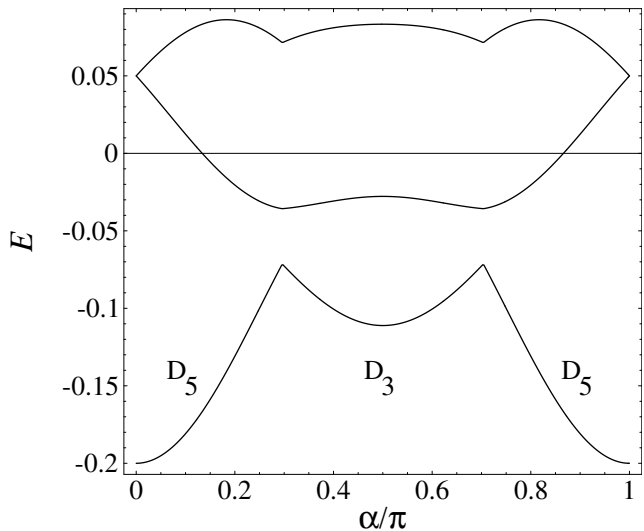,width=85mm,clip=}
}
\caption{\label{elecEigval:fig}
The splitting pattern of the electronic quintet in the $h\otimes H$ JT, as
a function of the mixing angle $\alpha$.
The lowest eigenvalue is nondegenerate, the two upper ones are both twofold
degenerate; energies are in units of $g^2\hbo$.
At each value of $\alpha$, $Q$ is taken at the optimal classical distortion
$Q_{\rm min}$.
}
\end{figure}

Figure~\ref{elecEigval:fig} shows the evolution of the quintet of
electronic eigenvalues as $\alpha$ takes its possible
values\footnote{
Values in the $\pi\leq\alpha\leq2\pi$ interval repeat those in the
$0\leq\alpha\leq\pi$ interval, except for a reversal of the associated
distortion $\hat{Q}_{\mu}\to-\hat{Q}_{\mu}$.
}.
Here, the molecular vibration operators are treated as classical variables,
and are taken at the relevant static JT minimum, of $D_5$ or $D_3$ symmetry,
appropriate for each value of $\alpha$.
Pure coupling of type 1 is recovered at $\alpha\!=\!0$ and $\pi$, while pure
coupling of type 2 occurs for $\alpha\!=\!\pi/2$.
At the special angles $\alpha_s=\arctan(3/\sqrt 5)\simeq 0.296\,\pi \simeq
53.3^\circ$ and $\pi\!-\!\alpha_s$, where cusps occur in the
$\alpha$-dependence of the eigenvalues, $D_5$ and $D_3$ minima are exactly
degenerate: the JT minima collapse into a single flat trough, and the
symmetry of the JT system is effectively larger than icosahedral
\cite{CeulemansII,Delos96,Paris97,noberry,Judd84,Judd99}.
Figure~\ref{elecEigval:fig} shows that for a given coupling energy
$g^2\hbo$, the lowering of the lowest electronic state (equal twice the
static JT energy gain) is strongest at $\alpha=0,\ \pi$ and weakest at
$\alpha_s,\ \pi\!-\!\alpha_2$.

To compute the exact vibronic spectrum of $\hat H$, full quantum-mechanical
treatment must be applied also to the vibrational operators.
The standard ladder-operator representation for the vibrational coordinates
$\hat{Q}_{\mu} =\left(\hat{b}_{\mu}^\dagger +\hat{b}_{\mu}\right)/\sqrt 2$
provides a natural product basis of harmonic phonon states centered on the
undistorted geometry, times the electronic quintet.
It is straightforward to write $\hat H$ matrix elements in this basis and
diagonalize the resulting matrix.
This basis centered on the high-symmetry point is especially convenient at
weak coupling, where converged low-energy vibronic eigenenergies and
eigenstates are obtained by including only few basis states.
For large $g$, the number of basis states needed for a given accuracy of
the low-energy levels grows rather quickly, approximately as $g^{2 n_{\rm
osc}}$, where $n_{\rm osc}=4$ for a $G$ mode and $n_{\rm osc}=5$ for an $H$
mode.
For this reason, and because of the extreme sparseness of this problem, the
Lanczos method of diagonalization \cite{Prelovsek00,Meyer89} is
especially suitable here.
We use a method very similar to the one employed in Ref.~\onlinecite{hbyh},
with the introduction of several refinements to deal with the problem of
Lanczos ghost states, and the symmetry analysis.
We check all results against basis truncation.  For most results, inclusion
of $n=35$ phonons in the basis is largely sufficient for an accuracy of
$10^{-6}\hbo$, but for occasional points in parameter space the basis
needs to include up to $n=60$ phonons (over 8 million states) to guarantee
the required accuracy.
The spectra computed with this method and reported below draw a full
quantitative link from weak through intermediate to strong coupling for the
$h\otimes G$ and $h\otimes H$ linear dynamical JT model.

\section{Vibronic spectra: energies}

\subsection{The $h\otimes G$ JT}

\begin{figure}
\centerline{
\epsfig{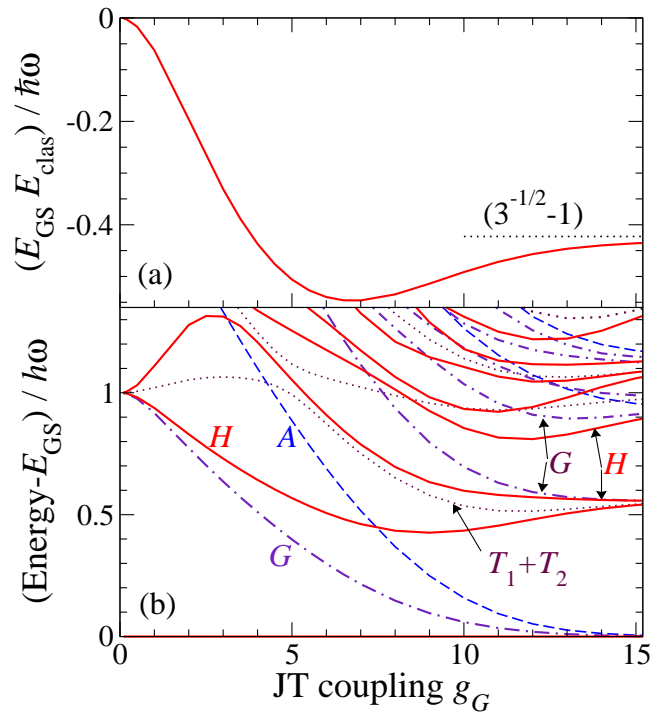}
}
\caption{\label{G-mode:fig}
Exact vibronic spectrum of a single $G$ mode coupled to a $h$ level, as a
function of the linear coupling parameter $g_G$.
(a) GS energy, to which the adiabatic minimum lowering $E_{\rm
clas}= \left(2 -\frac 1{18}g_G^2\right) \hbo$ is subtracted.
(b) Vibronic excitation energies (measured with respect to the GS).
All levels are labeled according to their symmetry as $A$ (dashed),
$T_1+T_2$ (dotted), $G$ (dot-dashed) and $H$ (solid).
}
\end{figure}

Figure~\ref{G-mode:fig}a shows the difference between the exact dynamical
JT GS energy and $E_{\rm clas}$ (the classical adiabatic energy
$-\frac1{18} g_G^2\hbo$ of the 10 $D_3$ wells plus the zero-point energy
$2\hbo$ of the uncoupled oscillators), as a function of the coupling
strength $g_G$ to a $G$ mode.
This residual energy represents the part of JT energy gain not ascribable
to trivial adiabatic lowering of the JT minima with respect to zero coupling:
%
%
this energy difference gauges quantum-mechanical effects on the vibrational
motion.
The rapid decrease in GS energy at weak coupling is due to the large drop
in quantum kinetic energy in the transition from harmonic oscillations
around one isolated adiabatic minimum to rapid tunneling among 10 shallow
minima.
As these minima move apart, tunneling is suppressed, and the plotted energy
difference turns back up to the strong-coupling value
$(3^{-1/2}-1)\hbo$,
which measures the zero-point energy gain due to phonon softening at the
anisotropic JT minima \cite{Lueders03,Leuven02}.
Indeed, the second-order expansion of the lowest APES around any of the
equivalent JT minima yields two normal modes (of symmetry $A_1$ and $A_2$
in the local $D_3$ group) both of frequency $\omega$, plus a
twofold-degenerate softer mode (of symmetry $E$) of frequency
$3^{-1/2}\omega$.
The difference in harmonic zero-point energy $2\, \frac12 \hbo + 2\,
\frac12 3^{-1/2}\hbo - 4\,\frac12 \hbo =
(3^{-1/2}-1)\hbo$ provides an accurate estimate for the exact vibronic
energy (Fig.~\ref{G-mode:fig}a) at strong coupling, where inter-well
tunneling is suppressed and the phonon dynamics reduces to harmonic
oscillations around the JT wells.
Note that the ``new'' frequencies at the wells are {\em independent of the
coupling strength $g_G$}, since for different coupling $g_G$, the set of
five APES differs only for a scaling factor, thus remaining self-similar.
At weak coupling, the harmonic frequencies at the JT wells remain
unchanged, but this expansion is practically irrelevant, since quantum
kinetic energy delocalizes the motion far enough from the minima for
anharmonic and nonadiabatic effects to dominate.

The vibronic GS symmetry remains $H$ at all couplings.
This is eventually a consequence of a Berry phase of $\pi$ acquired by the
electronic wavefunction as the distortions follow any of the cheapest
(5-corner) loops through adjacent $D_3$ minima: this entanglement makes the
totally symmetric $A$ vibronic combination of the wells less advantageous
than the $H$ combination, which thus remains the GS for all $g_G$.
Indeed, as seen in Fig.~\ref{G-mode:fig}b, even at strong coupling, the $A$
vibronic state lies above both the $H$ GS and a $G$ excitation
At strong coupling, these $5 (H) +4(G) +1 (A)$ states represent the 10
symmetry-adapted vibronic combinations of the harmonic GS's at the
10 $D_3$ wells.
Likewise, in the strong-coupling limit, higher excited states, represent
suitably symmetrized combinations of vibrational excitations in the wells.
In particular, the 20 states ($T_1+T_2+G+2\,H$) converging at strong
coupling to $3^{-1/2} \hbo$ represent the tunneling-symmetrized one-phonon
oscillations of the new softer frequency.
Similarly, the 20 states ($A+T_1+T_2+2\,G+H$) converging at strong coupling
to $\hbo$ represent the tunneling-symmetrized one-phonon oscillations of
the modes at the original frequency $\hbo$.

At weak coupling, the vibronic levels reconstruct the harmonic ladder, as
expected when $\hat{H}_{\rm e-v} $ is a weak perturbation.
The present exact calculation traces precisely the crossover from the weak
to strong coupling regime: for accurate numeric eigenvalues, see
Table~\ref{tabG} in the Appendix.
It is worth noting that the lowest $A$ level, which at strong coupling is a
low-energy tunneling excitation, correlates to the $2\,\hbo$ multiplet at
weak coupling \cite{Moate97}.
Finally, note that the $T_1$ and $T_2$ levels are exactly degenerate for
all values of the coupling.

\subsection{The $h\otimes H$ JT}

\begin{figure}
\centerline{
\epsfig{file=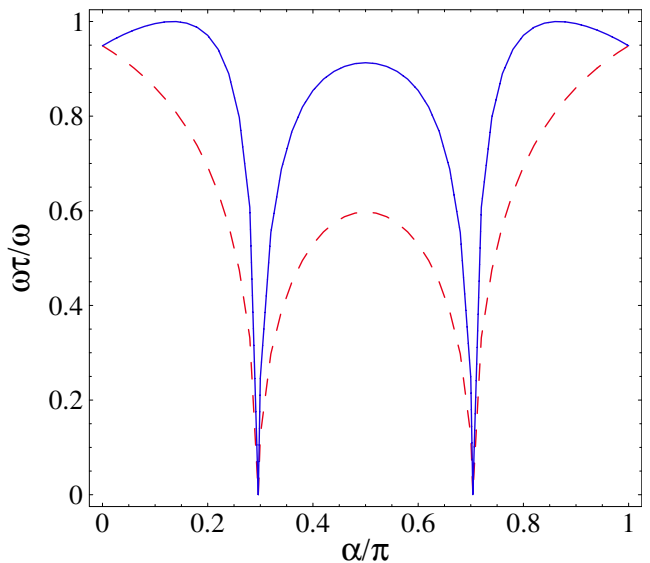,width=85mm,clip=}
}
\caption{\label{normalfreqAlpha:fig}
``Tangential'' normal frequencies $\omega_\tau$ of vibration around the JT
wells, in units of the original frequency $\omega$, as a function of
$\alpha$.
Both these modes are twofold degenerate.
As customary of linear JT, the frequency of the fifth (radial) normal mode
equals $\omega$, regardless of $\alpha$.
}
\end{figure}

We move on now to study the spectrum for a vibrational mode of $H$
symmetry.
The Hamiltonian $\hat H$ depends now on two parameters.
Each vibronic eigenvalue traces a 2-dimensional surface as $g$
and $\alpha$ (or equivalently $g_H^{(1)}$ and $g_H^{(2)}$) are varied.
The most effective  presentation of these spectra is obtained by means of
radial cuts at fixed $\alpha$, plus circular cuts at fixed $g$.

Like for the $G$ mode, the strong-coupling asymptotic vibronic energies,
are determined by the normal frequencies of oscillation around the
adiabatic JT minima, and these frequencies now depend on $\alpha$.
The five normal modes separate into a radial vibration of unchanged
frequency $\omega$ and symmetry $A_1$, plus two softer doubly-degenerate
(of $E_1$ and $E_2$ symmetry for $D_5$ wells and both of $E$ symmetry for
$D_3$ wells) tangential modes, represented in
Fig.~\ref{normalfreqAlpha:fig}.
For special $\alpha$ values, extra degeneracies occur:
\begin{itemize}
\item
for $\alpha=0$ and $\pi$, the two tangential modes become
degenerate at $\omega_\tau=(9/10)^{1/2} \omega$;
\item 
for $\alpha\simeq 0.134\,\pi$  
and $0.866\,\pi$,  
the upper tangential frequency tops at a maximum where it reaches
the frequency $\omega$ of the radial mode;
\item
at the special $\alpha=\alpha_s$ and $\pi-\alpha_s$ points, both tangential
frequencies vanish as the vibrations turn into free modes of pseudorotation
along the flat JT trough;
\item
in addition, for $\alpha=\pi/2$, both soft modes reach a local maximum, of
frequencies $\omega_\tau^{(1)}=(5/14)^{1/2} \omega$ and
$\omega_\tau^{(2)}=(5/6)^{1/2}\omega$.
\end{itemize}
Observe that the normal frequencies at the wells only depend on $\alpha$,
while they are {\em independent of the coupling strength $g$}, like for the
$G$ mode.
Again, these single-well harmonic frequencies determine the strong-coupling
asymptotic vibronic energies, but they are essentially irrelevant at weak
coupling, where inter-well tunneling dominates.

\begin{figure}
\centerline{
\epsfig{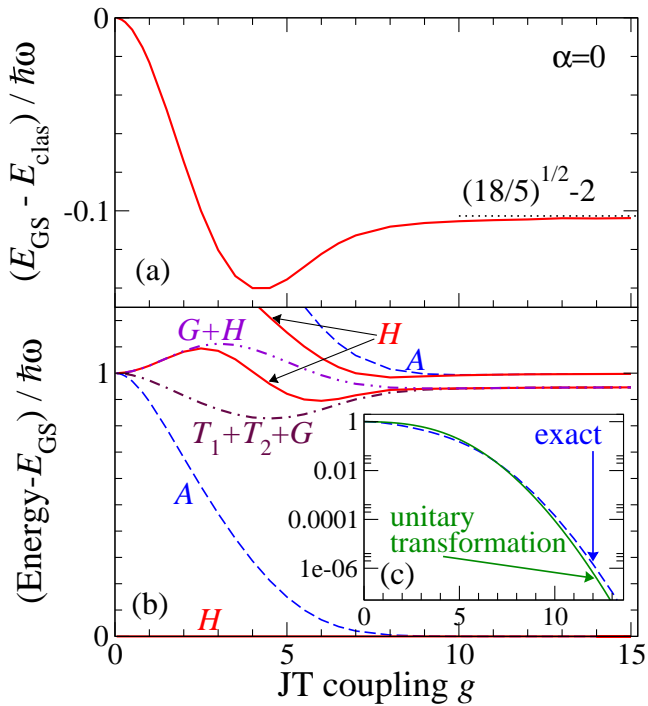}
}
\caption{\label{H-mode_theta0.0:fig}
Exact vibronic spectrum of a single $H$ mode coupled to a $h$ level, as a
function of the linear coupling parameter $g$, for $\alpha=0$.
(a) GS energy, to which the energy of the adiabatic JT wells
$E_{\rm clas}=\left(5/2 - g^2/10\right) \hbo$ is
subtracted.
(b) Vibronic excitation energies (measured with respect to the GS).
The levels are labeled according to their icosahedral symmetry even though,
for this special value of $\alpha$, the effective underlying symmetry is
spherical.
The conversion in spherical angular-momentum labels $[j]$ is $A\equiv [0]$,
$H\equiv [2]$, $G+H\equiv [3]$, $T_1+T_2+G\equiv[1]+[3]$.
(c) Comparison of the exact excitation energy of the lowest $A$ state (the
tunneling splitting) with the approximate expression obtained in
Ref.~\onlinecite{Moate97} by means of the unitary transformation method
\cite{Bates87}.
}
\end{figure}

Not unlike the $h\otimes G$ calculation, to draw the GS energy it is
convenient to subtract the quantity
\begin{equation} \label{eclas}
E_{\rm clas}(g,\alpha)=
\left\{\!\!
\begin{array}{ll}
\left[\frac 52 - \frac 1{10}(\cos\alpha\;g)^2\right] \hbo ,
	& 0\leq\alpha\leq\alpha_s \\
\\
\left[\frac 52 - \frac 1{18}(\sin\alpha\;g)^2\right] \hbo ,
	& \alpha_s\leq\alpha\leq\frac{\pi}2 \\
\end{array}\right.\!\!\!.
\end{equation}
$E_{\rm clas}(g,\alpha)$ represents the trivial zero-point energy $\frac 52
\hbo$ of $\hat{H}_{\rm vib}$ plus the adiabatic lowering of the $D_5$/$D_3$
JT well appropriate to that value of $\alpha$.

Figure~\ref{H-mode_theta0.0:fig}a reports the exact GS energy, for the
special mixing angle $\alpha=0$ (a mode of pure type $r=1$).
This energy difference converges rapidly to its asymptotic value
$[(18/5)^{1/2}-1]\hbo$ at strong-coupling.
Like for the $G$ mode, this is understood in terms of zero-point energy
associated to the four frequencies $\omega_\tau = (9/10)^{1/2}\omega$ plus
one frequency $\omega$ of the normal-mode vibrations at the JT minima.

The spectrum of excitations Fig.~\ref{H-mode_theta0.0:fig}b is much less
cluttered than that of the $G$ mode (Fig.~\ref{G-mode:fig}b).
The main reason is the larger than icosahedral effective symmetry of the
Hamiltonian at this specific coupling angle.
This larger (spherical) symmetry induces extra degeneracies among the
vibronic levels, which one could in fact label with spherical
angular-momentum labels.
The lowest excited state is of $A$ symmetry: it is the one state which at
strong coupling drops towards the $H$ GS.
Together, these $5 (H) +1 (A)$ states represent the 6 symmetrized
tunneling-split vibronic combinations of the harmonic GS's at the
6 $D_5$ wells.
An approximate analytical approach to this problem based on this idea (the
unitary transformation method \cite{Bates87}) yields
\cite{Moate97} an estimate of the excitation energy (solid line in
Fig.~\ref{H-mode_theta0.0:fig}c)\footnote{
The linear coupling parameters $k_\Lambda$ of Ref.~\onlinecite{Moate97} are
related to the dimensionless coupling parameters of the present work by
$k_{h_2}=\frac 12 g_H^{(1)} \hbo$,
$k_{h_1}=\frac 12 g_H^{(2)} \hbo$, and
$k_{g}=\frac 12 g_G \hbo$.
}: this approximation is seen to be fairly
accurate throughout the range of couplings.
The approximate expression underestimates the tunneling splitting at strong
coupling due to the simplifying assumption of isotropic JT wells, of
frequencies all equaling $\omega$, which therefore exaggerate the
localization of the distorted states.  Corrections due to the
anisotropy of the wells were introduced in Ref.~\onlinecite{Dunn03}.
In the Appendix, Table~\ref{tab0} reports a few of the vibronic
eigenenergies used to draw Fig.~\ref{H-mode_theta0.0:fig}.

Higher excitations cluster around the harmonic normal-mode energies at the
$D_5$ wells: the tangential $\hbo_\tau=(9/10)^{1/2}\hbo$ and the radial
$\hbo$.
The 24 states ($T_1 + T_2+2 G + 2 H$) around $\hbo_\tau$ are symmetrized
combinations of the four 1-phonon states per each of the six $D_5$ JT
wells.
Likewise, the 6 states ($H+A$) near $\hbo$ are combinations of the
1-radial-phonon states in the $6$ wells.
Note in particular that the two 1-phonon $H$ states remain almost
degenerate until $g\lesssim 2$, they split significantly in the interval
$2\lesssim g\lesssim 8$, and then re-converge again (to $\hbo_\tau$) at
strong coupling.
At strong coupling, higher excitations (not drawn) are found in the
overtone region $2\,\hbo_\tau$, and above.

\begin{figure}
\centerline{
\epsfig{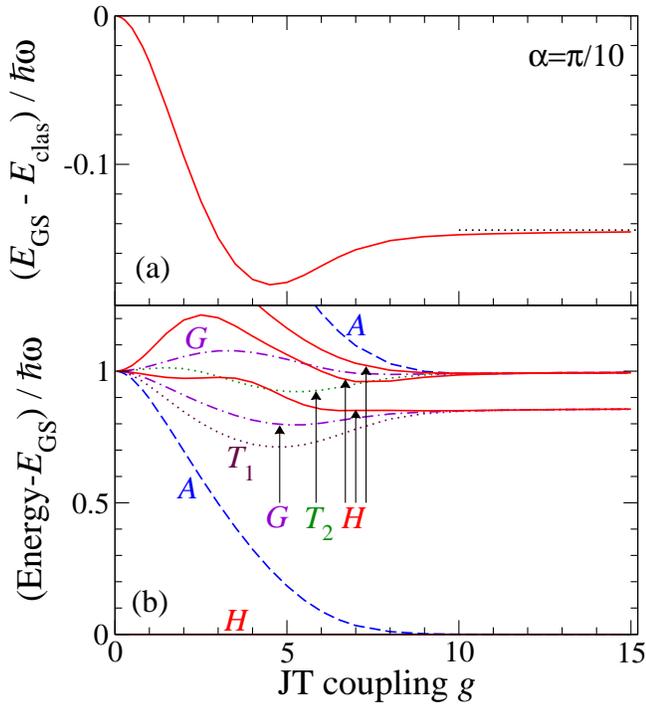}
}
\caption{\label{H-mode_theta0.314159:fig}
Exact vibronic spectrum of a single $H$ mode coupled to a $h$ level, as a
function of the coupling parameter $g$, for $\alpha=\pi/10$.
(a) GS energy, to which $E_{\rm clas}$ is subtracted.
(b) Vibronic excitation energies measured with respect to the $H$ GS.
All levels are labeled according to their symmetry as $A$ (dashed),
$T_{1\,/\,2}$ (dotted), $G$ (dot-dashed) and $H$ (solid).
}
\end{figure}

For $\alpha=\pi/10$, Fig.~\ref{H-mode_theta0.314159:fig}a shows the exact
GS energy referred to $E_{\rm clas}$.
This small value of $\alpha$ has little effect on the GS energy,
as compared to the $\alpha=0$ case.
On the contrary, the comparison of Fig.~\ref{H-mode_theta0.314159:fig}b
with Fig.~\ref{H-mode_theta0.0:fig}b shows that even such a small value of
$\alpha$ (i.e.\ a tiny admixture of the coupling of type $r=2$) is
sufficient to completely resolve the ``accidental'' degeneracies of the
$\alpha=0$ spectrum.
$A$ and $G$ states are little affected, while the
$T_1$/$T_2$ degeneracy is widely broken.
Likewise, the weak-coupling near degeneracy of the 1-phonon $H$ states for
$\alpha=0$ is now widely split.
Eventually, at strong coupling, these two $H$ states converge to different
tangential frequencies of the JT wells.

\begin{figure}
\centerline{
\epsfig{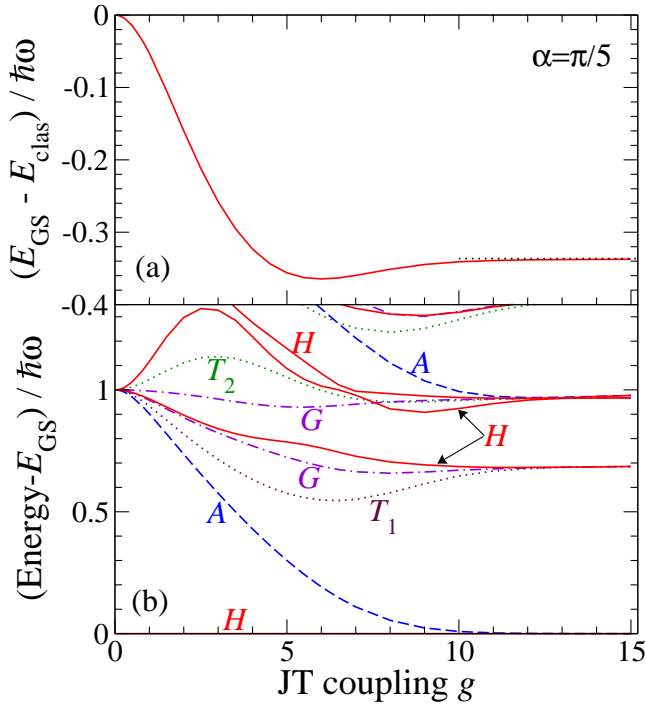}
}
\caption{\label{H-mode_theta0.628318:fig}
Same as Fig.~\ref{H-mode_theta0.314159:fig} for $\alpha=\pi/5$.
}
\end{figure}

Coming to the spectrum for $\alpha=\pi/5$,
Fig.~\ref{H-mode_theta0.628318:fig} shows a continuous evolution away from
$\alpha=0$ through $\alpha=\pi/10$.
We observe that, at strong coupling for both $\alpha=\pi/10$ and
$\alpha=\pi/5$, only one new frequency is seen in the strong-coupling
spectrum: that of decreasing $\omega_\tau$ for increasing $\alpha$
($\omega_\tau=0.86\,\omega$ for $\alpha=\pi/10$, $\omega_\tau=0.69\,\omega$
for $\alpha=\pi/5$).
Figure~\ref{normalfreqAlpha:fig} confirms that the frequency $\omega_\tau$
of the other tangential mode is very close to its maximum $\omega$, thus
very difficult to separate on this scale.
In practice all 18 relevant one-phonon states converge near $\hbo$ at
strong coupling.

\begin{figure}
\centerline{
\epsfig{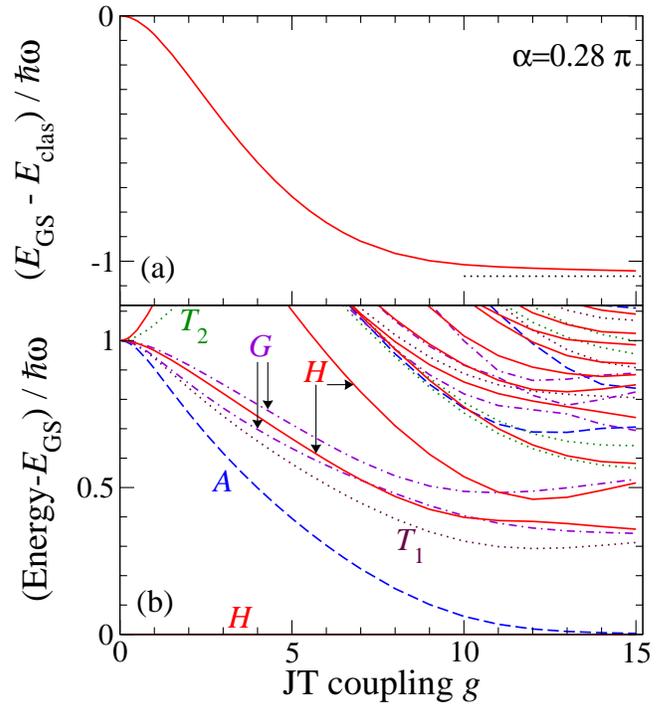}
}
\caption{\label{H-mode_theta0.879645:fig}
Same as Fig.~\ref{H-mode_theta0.314159:fig} for
$\alpha=0.28\,\pi\lesssim\alpha_s$.
}
\end{figure}

Moving further away from this accidental degeneracy, to $\alpha=0.28\,\pi$
(Fig.~\ref{H-mode_theta0.879645:fig}), the two new tangential frequencies
become $\omega_\tau^{(1)} \simeq 0.33\,\omega$ and
$\omega_\tau^{(2)} \simeq 0.61\,\omega$.
These frequencies are so small because of the nearness to the special angle
$\alpha_s$ where $D_5$ and $D_3$ minima degenerate to a flat trough (see
Fig.~\ref{normalfreqAlpha:fig}).
These small values explain the twofold reason why the spectrum of
Fig.~\ref{H-mode_theta0.879645:fig}b is so cluttered: (i) tunneling is only
weakly suppressed by the low inter-well barriers, and (ii) several
overtones and combinations of these soft tangential modes fit in the
considered energy range.
The large zero-point energy lowering shown in
Fig.~\ref{H-mode_theta0.879645:fig}a is also well accounted for by the
values of $\omega_\tau$.

\begin{figure}
\centerline{
\epsfig{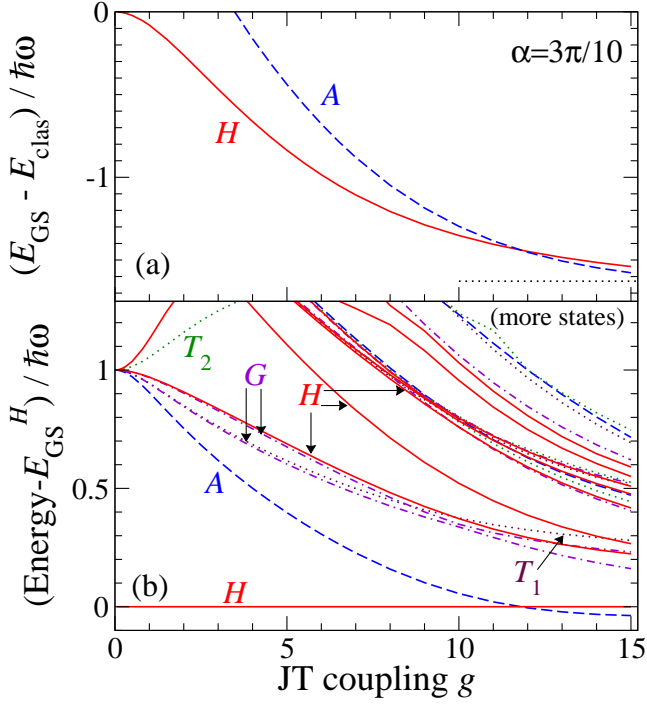}
}
\caption{\label{H-mode_theta0.942477:fig}
Exact vibronic spectrum of a single $H$ mode coupled to a $h$ level, as a
function of $g$, for $\alpha=0.3\,\pi\gtrsim\alpha_s$.
(a) GS energy, to which $E_{\rm clas}$ is subtracted; the GS
is of $H$ symmetry at weak coupling, and of $A$ symmetry at strong
coupling.
(b) Vibronic excitation energies measured with respect to the lowest $H$
state.
}
\end{figure}

The next point $\alpha=3\pi/10$ (Fig.~\ref{H-mode_theta0.942477:fig}) is
located immediately beyond $\alpha_s$, with $D_3$ minima barely prevailing
over $D_5$ distortions.
Here, the tangential frequencies are as low as $\omega_\tau^{(1)} \simeq
0.125\,\omega$ and $\omega_\tau^{(2)} \simeq 0.247\,\omega$: this accounts
for the huge zero-point energy lowering apparent in
Fig.~\ref{H-mode_theta0.942477:fig}a in the strong-coupling limit.
In contrast to all previous $\alpha$ values, where $D_5$ wells prevail, we
observe here the celebrated \cite{Delos96,Moate96,hbyh} level crossing
(occurring at a rather strong coupling $g\simeq 12$) to a nondegenerate
GS of $A$ symmetry.
The vibronic spectrum (Fig.~\ref{H-mode_theta0.942477:fig}b) is now even
more cluttered than for $\alpha=0.28\,\pi$, it shows an intricate pattern
of level crossings (states of different symmetries) and avoided crossings
(states of the same symmetry).
Several higher-lying levels have been omitted in the plot for clarity.
This is due to the exceedingly low normal frequencies at the wells, plus
the larger number of wells (10 instead of 6).
The clustering around the semiclassical frequencies is barely hinted, even
for the rather strong coupling at the right side of the plot.

\begin{figure}
\centerline{
\epsfig{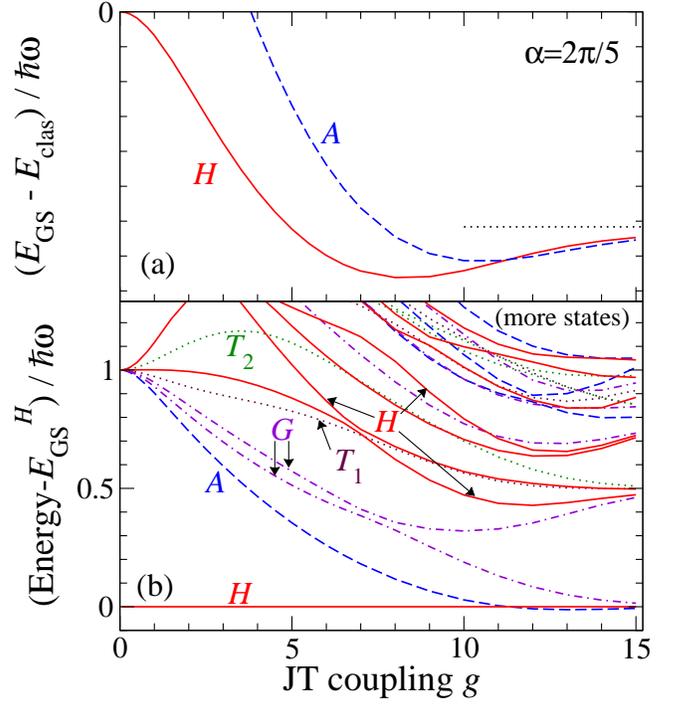}
}
\caption{\label{H-mode_theta1.256637:fig}
Same as Fig.~\ref{H-mode_theta0.942477:fig}, with $\alpha=2\pi/5$.
}
\end{figure}

\begin{figure}
\centerline{
\epsfig{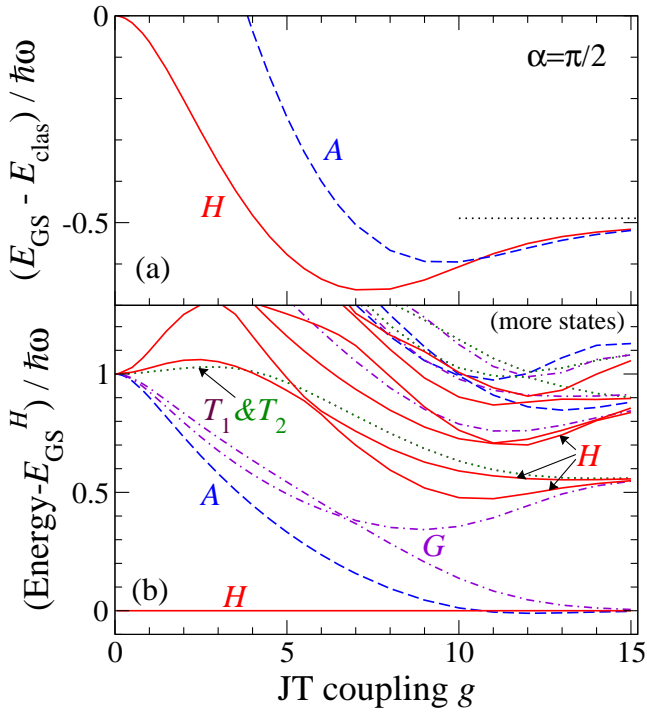}
}
\caption{\label{H-mode_theta1.570796:fig}
Same as Fig.~\ref{H-mode_theta0.942477:fig}, with $\alpha=\pi/2$.
}
\end{figure}

Coming now to $\alpha=2\pi/5$ (Fig.~\ref{H-mode_theta1.256637:fig}), the
tangential frequencies at the $D_3$ minima are now $\omega_\tau^{(1)}
\simeq 0.529\,\omega$ and $\omega_\tau^{(2)} \simeq 0.854\,\omega$, much 
larger than for the previous cut at $\alpha=3\pi/10$.
These values account for the reduced zero-point energy lowering shown in
Fig.~\ref{H-mode_theta1.256637:fig}a.
Also, the spectrum (Fig.~\ref{H-mode_theta1.256637:fig}b) is less dense:
the 10 tunneling states ($A+H+G$) converge to zero energy at strong
coupling; the 20 1-phonon tangential states converging to $\hbo_\tau^{(1)}$
are fairly well visible; the tangential $\hbo_\tau^{(2)}$, radial $\hbo$,
and overtone $2\,\hbo_\tau^{(1)}$ states are still substantially intermixed
even at the large coupling of the right side of
Fig.~\ref{H-mode_theta1.256637:fig}b.
The splittings between $T_1$ and $T_2$ pairs are consistently reduced here,
compared to previous values of $\alpha$.
Note in particular that, in contrast to the $\alpha$ range producing $D_5$
wells, both 1-phonon $H$ states correlate at strong coupling to the lowest
tangential mode of frequency $\hbo_\tau^{(1)}$.
A fairly general feature of the vibronic spectrum, especially visible in
Figs.~\ref{G-mode:fig}b, \ref{H-mode_theta1.256637:fig}b, and
\ref{H-mode_theta1.570796:fig}b, is the rapid drop of several 
vibronic energies, then followed by a successive climb back at strong
coupling where tunneling is suppressed and the motion gradually collapses
to oscillations around the adiabatic wells.

At $\alpha=\pi/2$ (Fig.~\ref{H-mode_theta1.570796:fig}), JT coupling of pure
$r=2$ type is acting.
As illustrated in Fig.~\ref{normalfreqAlpha:fig} above, the tangential
frequencies at the JT wells reach here a local maximum
$\omega_\tau^{(1)}\simeq 0.598\,\omega$ and $\omega_\tau^{(2)} \simeq
0.913\,\omega$.
Accordingly, the zero-point energy lowering of
Fig.~\ref{H-mode_theta1.570796:fig}a amounts to $[(5/14)^{1/2} +
(5/6)^{1/2}-2] \hbo= -0.490\,\hbo$.
The GS level crossing occurs now at a marginally smaller $g\simeq
10.7$, due to suppressed tunneling splitting.
The spectrum is now even less dense than for previous values of $\alpha$,
due to the larger frequencies at the JT wells, and the exact degeneracy of
the $T_1$ and $T_2$ pairs.
The 20 1-phonon tangential states ($G+2H+T_1+T_2$) converging to
$\hbo_\tau^{(1)}$ are well visible; the 20 states (of the same symmetries)
converging to $\hbo_\tau^{(2)}$ are still entangled with the 10 $\hbo$
radial states ($A+H+T_1+T_2$).

For $\pi/2<\alpha\leq\pi$ the spectra are mirror symmetric around
$\alpha=\pi/2$ to those reported for the interval $0\leq\alpha\leq \pi/2$.
The only difference is that all $T_1$ and $T_2$ states (which always come
in pairs) are exchanged.

\begin{figure}
\centerline{
\epsfig{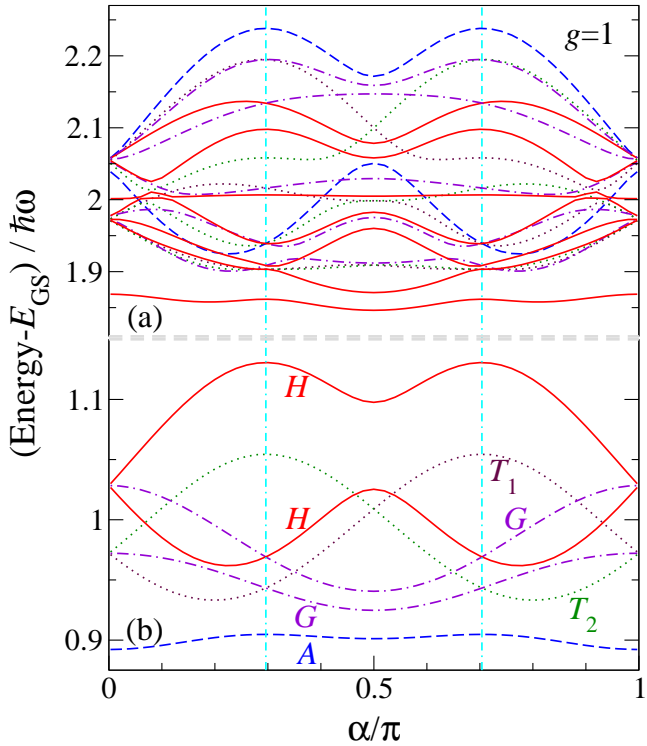}
}
\caption{\label{H-mode_g1:fig}
Exact vibronic spectrum of a single $H$ mode coupled to a $h$ electronic
quintet, as a function of $\alpha$, for $g=1$.
(a) The vibronic states related to the first overtone $2 \hbo$ region.
(b) The vibronic states perturbatively related to the ``fundamental''
excitation at $\hbo$.
}
\end{figure}

The $\alpha$-dependence of the low-lying excitations (measured with respect
to $E_{\rm GS}$) is best illustrated by fixed-$g$ circular cuts.
Figure~\ref{H-mode_g1:fig} reports the exact excitation energies at fixed
$g=1$ for the 25 vibronic states related to the 1-phonon excitation and for
the 75 states related to the 2-phonon overtone, as a function of $\alpha$.
This value of $g$ is sufficiently small for perturbation theory to provide
not especially bad estimates of the excitation energies drawn in
Fig.~\ref{H-mode_g1:fig}.
$A$, $G$, and $H$ energies are symmetric around $\alpha=\pi/2$, while $T_1$
and $T_2$ energies are antisymmetric functions of $\alpha$, coming in
pairs.
The lowest $A$ level is the lowest excitation for all $\alpha$: at strong
coupling it evolves into a low-energy tunneling partner with the $H$ state
(and a $G$ state when $D_3$ minima prevail).
The average position of the two $H$ states in the one-phonon multiplet is
located above $\hbo$: this causes the characteristic signature of JT in
molecular photoemission which was observed and discussed in
Ref.~\onlinecite{Manini03}.
Note that no singular behavior is observed near $\alpha_s$ and
$\pi-\alpha_s$ (vertical lines), in contrast to all quantities based on a
classical treatment of the vibrations, such as those plotted in
Figs.~\ref{elecEigval:fig} and \ref{normalfreqAlpha:fig}: the transition
from $D_5$ to $D_3$ wells is perfectly smooth in the quantum spectrum, and
signaled by only a number of extra degeneracies.

\begin{figure}
\centerline{
\epsfig{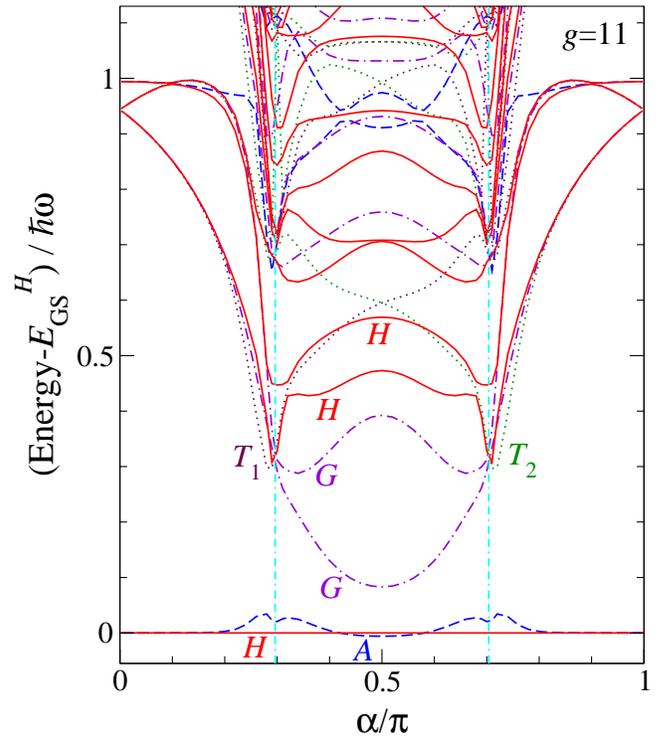}
}
\caption{\label{H-mode_g11:fig}
Low-energy region of the vibronic spectrum of a single $H$ mode coupled to
a $h$ level, as a function of $\alpha$, for $g=11$.
Energies are referred to the lowest $H$ state.
}
\end{figure}

Figure~\ref{H-mode_g11:fig} reports the excitation energies of several
low-lying vibronic states for $g=11$ (intermediate/strong coupling).
The tunneling partners are visible at the low-energy end, with the crossing
to an $A$ GS in the range $0.42\pi\lesssim\alpha\lesssim0.58\pi$.
The coupling is not strong enough to have the higher vibronic levels
organized according to the harmonic frequencies of
Fig.~\ref{normalfreqAlpha:fig} throughout the $\alpha$ range, but similar
trends are already recognizable: in particular for $\alpha\lesssim 0.2\pi$
and $\alpha\gtrsim 0.8\pi$, the vibronic levels follow quite closely the
harmonic frequencies of Fig.~\ref{normalfreqAlpha:fig}, and their overtones
and combinations.
As observed in the discussion of Fig.~\ref{H-mode_theta0.942477:fig},
the vibronic spectrum in the strong tunneling regions at and around the
flat-trough values $\alpha=\alpha_s$ and $\pi-\alpha_s$ is extremely
congested: only the lowest-lying states are drawn in figure.
Accordingly, most vibronic levels undergo a fast movement in that region,
but always evolve continuously across $\alpha=\alpha_s$ and $\pi-\alpha_s$.
In the limit of infinitely large $g$, the singular behavior of
Fig.~\ref{normalfreqAlpha:fig} would eventually emerge even in the exact
quantum spectrum.
The very large slope in the $\alpha$ dependence of many vibronic energies
may prove useful in future precise determinations of the JT coupling
parameters from spectroscopic data.

\section{Photoemission spectra}

Several spectral properties of experimental interest are readily accessed
by the exact diagonalization method at hand.
Photoemission spectra (PES) involving the quintet $h$ molecular orbital are
of course affected by JT coupling to the degenerate vibrations.
If, as is usually the case, the photoemitted electron kinetic energy is
much larger than $\hbo$, photoemission occurs in a time much shorter than
that characteristic of phonon dynamics.
In this limit, it is convenient to apply the {\em sudden
approximation} \cite{Cederbaum77,Koppel84}, where the photoemission process
is described simply by the operator $\hat{c}_{m\sigma}$ suddenly destroying
a spin-$\sigma$ electron in orbital component $m$ of the quintet level.
The phonon shakeup contributions to the PES are obtained in terms of the
matrix elements of this hole-creation operator $\hat{c}_{m \sigma}$ between
the initial configuration and the final vibronic states.
If we assume that the initial temperature is negligible ($k_{\rm
B}T\ll\hbo$, only the GS $|{\rm GS}\rangle$ is initially
populated), then the PES intensity for the creation of a spin-$\sigma$ hole
in orbital $m$ is
\begin{equation}\label{FermiGR}
I_{m\sigma}(E) \propto
\sum_{f}
\left|\langle f| \hat{c}_{m \sigma}
|{\rm GS}\rangle \right|^2
\, \delta\!\left(E - E_{f}\right) \,,
\end{equation}
where averaging over the components is implied whenever $|{\rm GS}\rangle$
is degenerate.
If the individual contributions of different orbital and spin component are
not separate, the total PES intensity is then
\begin{equation}\label{totalPES}
 I(E) = \frac 1{10} \sum_{m\sigma} I_{m\sigma}(E)\,.
\end{equation}

For generic occupancy of the $h$ level, the spectrum is generally affected
not only by JT, but also by intra-shell electron-electron repulsion, and
the ensuing multiplet structure \cite{Qiu01,Lueders02,Lueders03}:
calculation of the PES in such conditions must take into account several
other system-specific interactions.
It thus goes beyond the study of the single-mode rather general model
considered in this work.
However, electron-electron repulsion plays no role in two relevant cases,
which are therefore completely described by the JT Hamiltonian $\hat H$ of
Eqs.~\eqref{modelhamiltonian}-\eqref{JT-hamiltonian}: (i) when a {\em
single electron} occupies the quintet orbital {\em in the initial state},
and (ii) when the orbital quintet is initially completely full and
photoemission produces a {\em single hole in the final state}.
In the first case, JT only affects the initial configuration, while in the
second case, JT only affects the final states.
The application of exact diagonalization to the two cases separately
provides spectra where all non-adiabatic effects and phenomena neglected in
the Frank-Condon approximation \cite{Mahapatra99} are fully included.

\subsection{JT in the initial states}
\label{JTinitial}

\begin{figure}
\centerline{
\epsfig{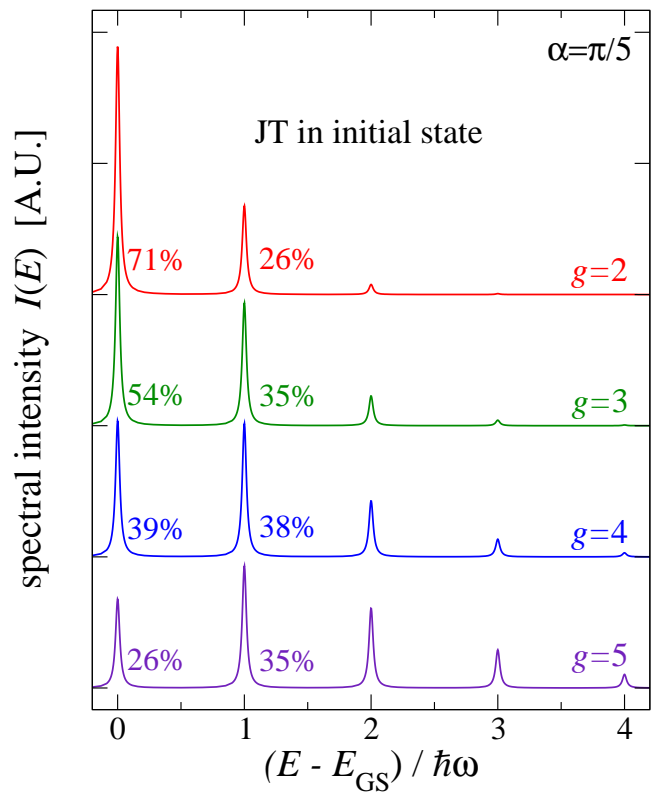}
}
\caption{\label{H-JTinPES_theta0.628318:fig}
Zero-temperature PES obtained by fast removal of the single electron from a
$h$ orbital coupled to a single $H$ mode, for several values of the
coupling $g$, and $\alpha=\pi/5$.
Energies are referred to the GS; a phenomenological broadening
HWHM=$0.02\,\hbo$ provides a finite width to the peaks.
Fractional spectral weights of the 0-phonon and 1-phonon line are listed.
}
\end{figure}

When JT only affects the initial states, the molecule is instantaneously
brought from the multi-well intrinsically nonadiabatic vibronic GS to a
linear superposition of the final harmonic oscillator states.
According to Eqs.~\eqref{FermiGR} and \eqref{totalPES}, the peaks in the
spectrum are located at the final-state energies $E_f$, which here simply
involve single or multiple excitations of the harmonic phonon $\hbo$.
The spectrum is therefore a sequence of regularly spaced peaks, with
intensity of the 0, 1, 2...\ -phonon line proportional to the square
modulus of the corresponding total component in the initial vibronic GS.

Figure~\ref{H-JTinPES_theta0.628318:fig} presents precisely a spectrum with
this simple structure, for an $H$ mode characterized by $\alpha=\pi/5$.
According to the standard electron-phonon (non-JT) Frank-Condon picture,
the percentage of spectral intensity displaced away from the 0-phonon to
the multi-phonon excitations increases with the distance between the
initial and final equilibrium configurations, which, in turn, is
proportional to the linear coupling parameter $g$.

\begin{figure}
\centerline{
\epsfig{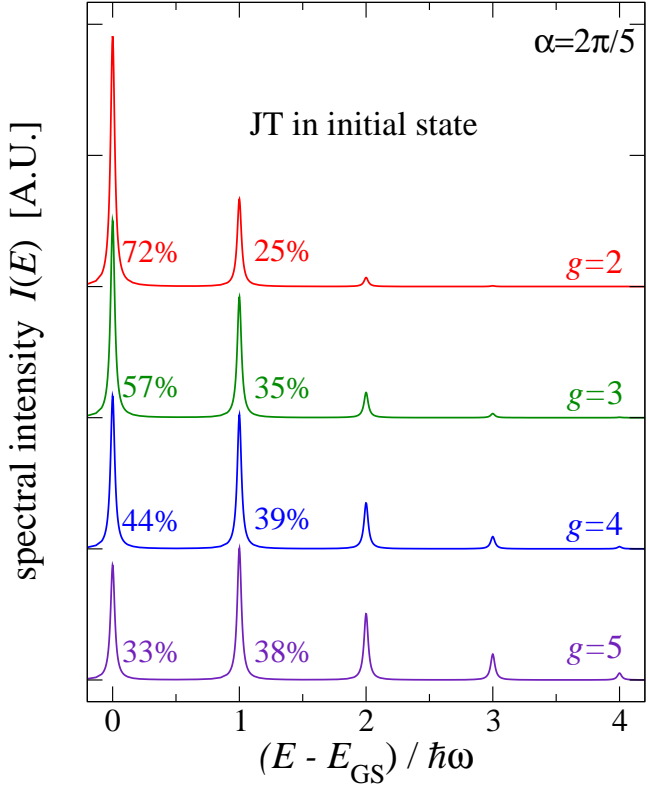}
}
\caption{\label{H-JTinPES_theta1.256637:fig}
Same as Fig.~\ref{H-JTinPES_theta0.628318:fig} for  $\alpha=2\pi/5$.
}
\end{figure}

The situation is qualitatively similar for $\alpha=2 \pi/5$ (and for the
case of a $G$ mode, which is omitted for brevity).
The only quantitative difference to observe in
Fig.~\ref{H-JTinPES_theta1.256637:fig} compared to
Fig.~\ref{H-JTinPES_theta0.628318:fig} is that for $\alpha=2 \pi/5$
spectral weight is preferentially transfered to the 1-phonon line, while
for $\alpha=\pi/5$ it is more equally distributed among several phonon
excitations.

The spectra in
Figs.~\ref{H-JTinPES_theta0.628318:fig}-\ref{H-JTinPES_theta1.256637:fig}
are obtained by means of a complete Lanczos calculation of the GS
wavefunction.
This is straightforward, as long as zero temperature is addressed as in the
present work.
If instead finite-temperature was to be considered, many initial vibronic
states would be required for a thermal average, and this is generally a
difficult task for the Lanczos method \cite{Bordoni04}: completely
different techniques, e.g.\ based on Monte Carlo, could prove more
convenient there.

\subsection{JT in the final states}
\label{JTfinal}

\begin{figure}
\centerline{
\epsfig{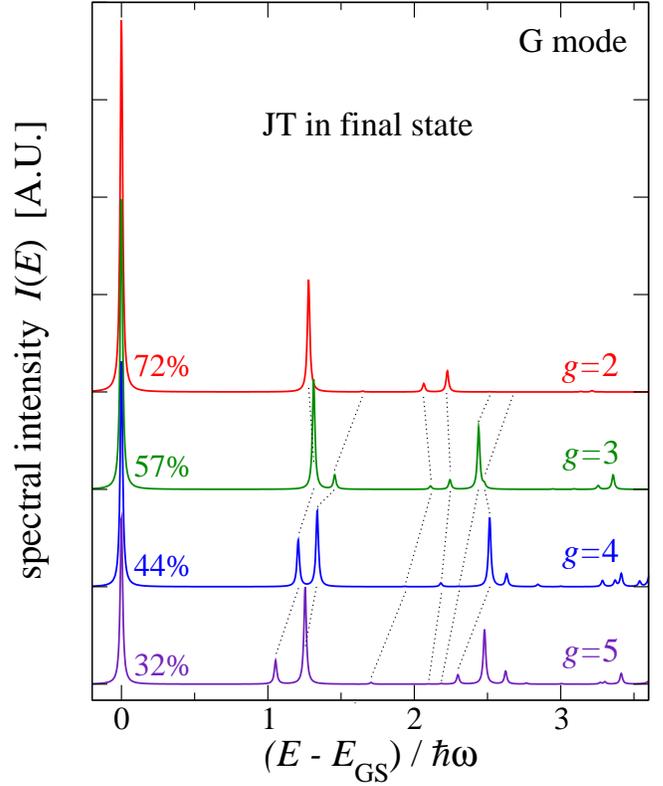}
}
\caption{\label{G-modePES:fig}
Zero-temperature PES obtained by fast removal of an electron from a filled
$h$ orbital coupled to a single $G$ mode, for several values of the
coupling $g$.
Energies are referred to the GS; a phenomenological broadening
HWHM=$0.01\,\hbo$ provides a finite width to the peaks.
Fractional spectral weights of the ``zero-phonon'' line are listed.
}
\end{figure}

When JT affects the final states, the energies $E_{f}$ marking the PES
peaks are the vibronic energies of
Figs.~\ref{G-mode:fig},~\ref{H-mode_theta0.0:fig}-\ref{H-mode_g11:fig}.
The transparent structure of a harmonic spectrum is now lost.
However, the calculation of the PES is straightforward, as $|{\rm
GS}\rangle$ is now the 0-phonon state of the filled-shell non-JT
molecule
\footnote{
Computationally, the Lanczos method is even more convenient here than when
JT occurs in the initial state.
Indeed, provided that the starting vector of the Lanczos chain is
$\hat{c}_{m\sigma} | {\rm GS}\rangle$ \cite{Prelovsek00}, the tridiagonal
Lanczos representation of $\hat H$ yields automatically the matrix elements
$\langle f| \hat{c}_{m \sigma} | {\rm 0}\rangle$, without any need to store
any Lanczos vector and reconstruct any of the actual eigenstates of the
original matrix.
All reported spectra are converged with respect to the finite length of the
Lanczos chain.
}.
The symmetry of the state $\hat{c}_{m\sigma} |{\rm GS}\rangle$ occurring in the
matrix element of Eq.~\eqref{FermiGR} is that of the electronic operator
$\hat{c}_{m\sigma}$, namely $H$.
As a consequence, only transitions to $H$ final states $\langle f|$ are
possible, while matrix elements to all other vibronic states vanish by
symmetry.
Peaks occur therefore only at the energies marked by the solid lines in
Figs.~\ref{G-mode:fig},~\ref{H-mode_theta0.0:fig}-\ref{H-mode_g11:fig}.

\begin{figure}
\centerline{
\epsfig{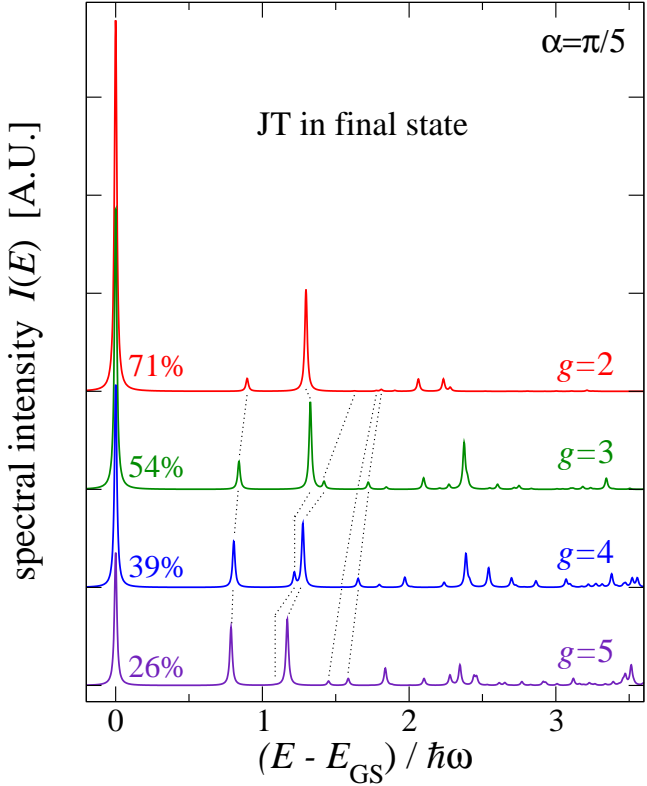}
}
\caption{\label{H-modePES_theta0.628318:fig}
Zero-temperature PES obtained by fast removal of an electron from a filled
$h$ orbital coupled to a single $H$ mode, for several values of the
coupling $g$ and, $\alpha=\pi/5$.
Energies are referred to the GS; a phenomenological broadening
HWHM=$0.01\,\hbo$ provides a finite width to the peaks.
Fractional spectral weights of the ``zero-phonon'' line are listed.
}
\end{figure}

Figure~\ref{G-modePES:fig} reports the computed PES for a coupled $G$ mode
and a few intermediate values of $g$.
The positions of the low-lying peaks correspond to the solid lines of
Fig.~\ref{G-mode:fig}b, and are readily followed.
Strong intensity transfers due to level crossings are well visible.
The intensity of the ``zero-phonon'' line (better referred to as the
vibronic GS) decreases steadily as $g$ increases \footnote{
The zero-phonon line intensity is the same in the initial and final JT
configurations (e.g.\ in Figs.~\ref{H-JTinPES_theta0.628318:fig} and
\ref{H-modePES_theta0.628318:fig}), since the involved matrix elements are
the same.
}: the intensity
lost there redistributes across an intricate vibronic spectrum.

\begin{figure}
\centerline{
\epsfig{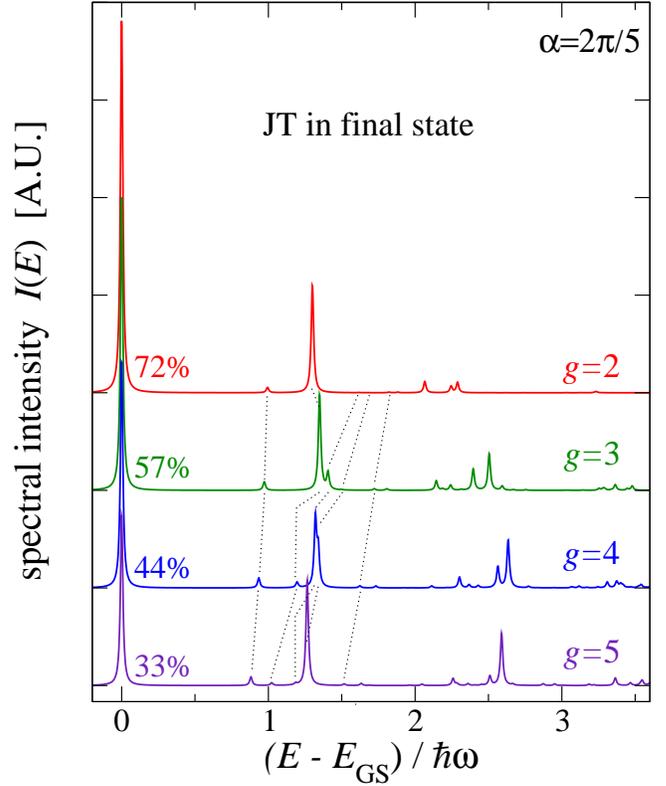}
}
\caption{\label{H-modePES_theta1.256637:fig}
Same as Fig.~\ref{H-modePES_theta0.628318:fig} for  $\alpha=2\pi/5$.
}
\end{figure}

Figures~\ref{H-modePES_theta0.628318:fig} and
\ref{H-modePES_theta1.256637:fig} report the computed PES for a
coupled $H$ mode characterized by the same value of $\alpha$ of
Fig.~\ref{H-mode_theta0.628318:fig} ($\pi/5$, producing $D_5$ minima) and
of Fig.~\ref{H-mode_theta1.256637:fig} ($2\pi/5$, producing $D_3$ minima)
respectively.
The positions and intensities of several low-lying peaks can be traced:
several intensity transfers due to level crossings are well visible.
Again, the intensity of the ``zero-phonon'' line decreases consistently as
$g$ is increased, while the intensity lost there redistributes across a
rather dense spectrum of vibronic levels.

Figures ~\ref{G-modePES:fig}-\ref{H-modePES_theta1.256637:fig} illustrate
in detail the general finding \cite{Manini03} that in the PES of dynamic JT
system with weak to intermediate coupling, the ``1-phonon line'' splits
into essentially two features, usually with most spectral weight in the
upper peak, blue-shifted up to 30\% higher energy than $\hbo$.
%
%
The blue shift of the 1-phonon satellite is not specific to the $h\otimes
(G+H)$ JT model.
For example, similar conclusions apply to the $e\otimes E$ problem
\cite{Bersuker}.  Indeed a 15\% blue shift of the first vibronic
satellite is clearly observable in the PE spectra of benzene
\cite{Baltzer97etal}.

A general feature of the displayed PES is that the perturbative structure
of $n$-phonon states, still fairly well recognizable for $g=2$), is
rapidly lost in the PES for $g\gtrsim 3$.
The only other general consideration we can draw from these spectra is that
not much regularity is to be expected: unless coupling is very weak, phonon
shakeups in the PES from a JT molecule build very intricate spectral
structures, which are strongly $g$-dependent (also $\alpha$-dependent, for
$H$ modes), even for a single phonon mode.

\section{Discussion}

Understanding the general features of a vibronic spectrum is a necessary
step towards the interpretation and assignment of the spectra of molecules
and molecular ions.
This has been undertaken and carried out quite extensively for electronic
doublets, triplets \cite{Bersuker,Englman,Martinelli91etGrosso} and more
complex electronic configurations \cite{Qiu01,Sookhun03,Li03}.
The present paper fills a gap in the theory of icosahedral systems,
providing general guidelines for the interpretation of vibronic PES of the
electronic quintet in the $h\otimes (G+H)$, restricting to the
idealization of a single mode.
In practical applications, the situation is much more intricate.
All actual icosahedral systems are many-mode systems, where the inter-mode
interaction generates a much less coherent spectrum than simple non-JT
phonons \cite{Gattari03}.
The present theory allows the basic understanding of the PES of icosahedral
systems wherever the coupling to a single high-energy mode is dominant,
the other couplings only contributing to the background.
The couplings of fullerene C$_{60}$ are approximately of this kind, and the
spectral of Sect.~\ref{JTfinal} allow to rationalize some observed PES
features in terms of an effective single-mode theory.

On the other hand, the spectra of Sect.~\ref{JTinitial} describe a situation
where a single electron in a $h$ state is rapidly removed.
This cannot occur in practically accessible ionic states of C$_{60}$, but
could well occur in singly charged negative ions of higher icosahedral
fullerenes, under the condition that the LUMO is a quintet.
By particle-hole symmetry, the same theory applies to {\rm inverse} PES,
where a single hole in a quintet is filled.
This is relevant, e.g.\ for inverse PES of C$_{60}^+$.

The difficulty of the many-modes JT \cite{Bersuker,ManyModes} is that no
linear superposition of the spectra of single modes occurs (except for the
weak-coupling limit).
This means that any quantitative description of the dynamics requires to
include all modes together in a single calculation.
The method of Lanczos exact diagonalization used in the present work is by
no means limited to the single-mode case.
Indeed, Ref.~\onlinecite{Manini03} deals successfully with the intricacies
of the C$_{60}$ many-mode problem, also accounting for thermal effects.
The same theory used in Sect.~\ref{JTfinal} for PES could be applied to an
optical transition from a nondegenerate electronic $a$ state to a final $h$
state, if that dipole-forbidden transition could be realized.
Modifications of the same method could be employed to study phonon shakeups
in optical absorption/emission in dipole allowed transitions involving
electronic degenerate states at both sides: these are often associated to
``product'' JT systems \cite{Qiu01,Ceulemans00}, and with due account of
``term'' exchange interactions, JT would account for the vibronic
``decorations'' of the spectra, not unlike PES.
More applications of the Lanczos method (including generalizations thereof
\cite{Meyer89}) to other spectroscopical applications are therefore
expected in the future.

\section*{Acknowledgments}

The author is indebted C.\ A.\ Bates, A.\ Bordoni, P.\ Gattari, I.\ D.\
Hands, and E.\ Tosatti for useful discussion.
This work was partly supported by FIRB RBAU017S8R operated by INFM.

\appendix
\section{Exact eigenenergies}
\label{app}

\begin{table}[]
\begin{center}
\begin{tabular}{ccccccc}
\hline
\hline
	&	1	&	2	&	5	&	10	&	15\\
\hline
$A$	&	      3.724144	&	      3.163979	&	      0.990680	&	     -3.888699	&	    -10.881188\\
$A$	&	      5.094787	&	      4.788557	&	      2.428383	&	     -2.782559	&	     -9.799588\\
$T_{1/2}$&	      2.903273	&	      2.634575	&	      1.098218	&	     -3.511597	&	    -10.336507\\
$T_{1/2}$&	      3.745236	&	      3.221392	&	      1.222379	&	     -3.110401	&	     -9.893907\\
$G$	&	      2.799066	&	      2.355251	&	      0.503564	&	     -3.987637	&	    -10.885576\\
$G$	&	      3.859923	&	      3.502678	&	      1.672126	&	     -3.351358	&	    -10.317040\\
$H$	&	      1.882022	&	      1.581814	&	      0.105229	&	     -4.046978	&	    -10.887960\\
$H$	&	      2.819443	&	      2.411645	&	      0.673159	&	     -3.612510	&	    -10.340670\\
$H$	&	      2.992497	&	      2.859999	&	      1.158021	&	     -3.449058	&	    -10.317965\\
$H$	&	      3.746243	&	      3.228490	&	      1.359820	&	     -3.191928	&	     -9.925392\\
$H$	&	      3.793559	&	      3.345318	&	      1.504876	&	     -3.113374	&	     -9.815484\\
$H$	&	      3.888832	&	      3.646643	&	      1.810633	&	     -2.941030	&	     -9.781893\\
\hline
\hline
\end{tabular}
\end{center}
\caption{\label{tabG}
Exact eigenenergies of several low-lying vibronic states of $h\otimes G$
linear JT for selected values of the coupling parameter $g$.  Energies are in
units of $\hbo$.
}
\end{table}

\begin{table}
\begin{center}
\begin{tabular}{cccccc}
\hline
\hline
$g$	&	1	&	2	&	5	&	10	&	15	\\
\hline
$A$	&	      3.269039	&	      2.704296	&	      0.014206	&	     -7.605244	&	    -20.103759	\\
$A$	&	      4.416117	&	      4.031416	&	      1.213521	&	     -6.611668	&	    -19.106543	\\
$T_{1/2}$&	      3.349001	&	      2.940441	&	      0.696237	&	     -6.663875	&	    -19.157685	\\
$G$	&	      3.349001	&	      2.940441	&	      0.696237	&	     -6.663875	&	    -19.157685	\\
$G$	&	      3.405264	&	      3.109744	&	      0.907404	&	     -6.663079	&	    -19.157685	\\
$H$	&	      2.376848	&	      2.025472	&	     -0.135556	&	     -7.605387	&	    -20.103759	\\
$H$	&	      3.405223	&	      3.107032	&	      0.786076	&	     -6.663301	&	    -19.157685	\\
$H$	&	      3.405264	&	      3.109744	&	      0.907404	&	     -6.663079	&	    -19.157685	\\
$H$	&	      4.245489	&	      3.651009	&	      1.018708	&	     -6.613430	&	    -19.106543	\\
\hline
\hline
\end{tabular}
\end{center}
\caption{\label{tab0}
Exact eigenenergies of several low-lying vibronic states of $h\otimes H$
linear JT for $\alpha=0$ and selected values of the coupling intensity $g$.
Energies are in units of $\hbo$.
}
\end{table}

\begin{table}
\begin{center}
\begin{tabular}{cccccc}
\hline
\hline
$g$	&	1	&	2	&	5	&	10	&	15	\\
\hline
$A$	&	      3.286920	&	      2.838680	&	      1.066520	&	     -3.082956	&	     -9.369654	\\
$A$	&	      4.309107	&	      3.872518	&	      2.044770	&	     -2.121879	&	     -8.473972	\\
$T_1$	&	      3.318833	&	      2.915855	&	      1.247562	&	     -2.696110	&	     -8.854724	\\
$T_1$	&	      4.287826	&	      3.838287	&	      2.068227	&	     -1.958552	&	     -8.086429	\\
$T_2$	&	      3.434328	&	      3.237084	&	      1.880242	&	     -2.280608	&	     -8.537845	\\
$T_2$	&	      4.285742	&	      3.829377	&	      2.016910	&	     -2.178646	&	     -8.345314	\\
$G$	&	      3.333101	&	      2.956614	&	      1.370430	&	     -2.614428	&	     -8.851809	\\
$G$	&	      3.364209	&	      3.029490	&	      1.485305	&	     -2.346950	&	     -8.538216	\\
$H$	&	      2.382792	&	      2.089510	&	      0.694933	&	     -3.124845	&	     -9.370196	\\
$H$	&	      3.345469	&	      2.978750	&	      1.375559	&	     -2.572543	&	     -8.849910	\\
$H$	&	      3.510110	&	      3.423708	&	      1.778880	&	     -2.340165	&	     -8.537793	\\
$H$	&	      4.242742	&	      3.729200	&	      1.947019	&	     -2.305749	&	     -8.491719	\\
\hline
\hline
\end{tabular}
\end{center}
\caption{\label{tabpi4}
Same as Table~\ref{tab0}, for $\alpha=\pi/4$.
}
\end{table}

\begin{table}
\begin{center}
\begin{tabular}{cccccc}
\hline
\hline
$g$	&	1	&	2	&	5	&	10	&	15	\\
\hline
$A$	&	      3.282422	&	      2.807794	&	      0.866002	&	     -3.650697	&	    -10.518971	\\
$A$	&	      4.431838	&	      4.165069	&	      2.165531	&	     -2.669479	&	    \ -9.635062	\\
$T_{1/2}$	&	      3.390187	&	      3.098692	&	      1.498544	&	     -3.029840	&	    \ -9.956428	\\
$G$	&	      3.305993	&	      2.869310	&	      1.027386	&	     -3.525390	&	    -10.511108	\\
$G$	&	      3.321806	&	      2.908694	&	      1.077499	&	     -3.306272	&	    \ -9.969104	\\
$H$	&	      2.381251	&	      2.074260	&	      0.534055	&	     -3.662262	&	    -10.516120	\\
$H$	&	      3.406622	&	      3.132388	&	      1.463674	&	     -3.184920	&	    \ -9.966811	\\
$H$	&	      3.478854	&	      3.325880	&	      1.517402	&	     -3.071858	&	    \ -9.959172	\\
$H$	&	      4.227504	&	      3.672462	&	      1.693740	&	     -2.935829	&	    \ -9.678823	\\
$H$	&	      4.252143	&	      3.734125	&	      1.787623	&	     -2.900136	&	    \ -9.660194	\\
\hline
\hline
\end{tabular}
\end{center}
\caption{\label{tabpihalf}
Same as Table~\ref{tab0}, for $\alpha=\pi/2$.
}
\end{table}

Tables~\ref{tabG}-\ref{tabpihalf} collect a number of exact vibronic
eigenenergies of the orbital-quintet single-mode linear JT model.
Note that $E_{\rm clas}$ was {\em not} subtracted here, therefore the
tabulated energies represent precisely the eigenvalues of the Hamiltonian
as defined in Eq.~\eqref{modelhamiltonian}.
All reported digits are significant.
Extra degeneracies are apparent in all tables except Table~\ref{tabpi4}
($\alpha=\pi/4$).
However, the apparent degeneracy between the $A$ states and $H$ states for
$g=15$, $\alpha=0$ (Table~\ref{tab0}) only indicates that the (tunneling)
splitting between these states is smaller than the energy resolution of the
diagonalization (see also Fig.~\ref{H-mode_theta0.0:fig}).



\end{document}